\documentclass[preprint]{aastex}
\usepackage{mathptmx}
\usepackage{graphicx}
\usepackage{epstopdf}
\usepackage{pdflscape}
\begin{document}

\title{Space Weathering Trends Among Carbonaceous Asteroids}
\author{Kaluna, H. M.$^1$, Masiero, J. R. $^2$, Meech, K. J.,$^1$}
\affil{$^1$Institute for Astronomy, University of Hawaii, 2680 Woodlawn Dr., Honolulu-HI-96822, USA}
\affil{$^2$Jet Propulsion Laboratory/Caltech,Pasadena, CA, USA}
\email{kaluna@hawaii.edu, Joseph.Masiero@jpl.nasa.gov, meech@ifa.hawaii.edu}

\begin{abstract}
We present visible spectroscopic
and albedo data of the 
2.3 Gyr old Themis family and the $<$10~Myr old Beagle sub-family.  
The slope and albedo variations between these two families indicate C-complex 
asteroids become redder and darker in response to space weathering. 
Our observations of Themis family members confirm previously observed
trends where phyllosilicate absorption features are less common among
small diameter objects.  Similar trends in the albedos of large ($>$15~km) and 
small ($\le$15~km) Themis members suggest these phyllosilicate feature and albedo 
trends result from regolith variations as a function of diameter.  Observations
of the Beagle asteroids show a small, but notable fraction of members with phyllosilicate features.
The presence of phyllosilicates and the dynamical association of the 
main-belt comet 133P/Elst-Pizarro with the Beagle 
family imply the Beagle parent body was a 
heterogenous mixture of ice and aqueously altered minerals.
\end{abstract}


\section {Introduction}
Space weathering studies on asteroids have primarily focused on the alteration
of the silicate rich S-complex asteroids. These 
moderate albedo \cite[p$_v \sim$0.22;][]{Mainzer:2011} asteroids are known 
to spectrally darken, redden and have increasingly suppressed absorption bands as a 
function of time \citep{Belton:1992,Belton:1994,Binzel:1996,Chapman:1996}.  
These spectral changes are attributed to the vapor deposition
of sub-micron metallic iron (SMFe) particles onto grains
during micrometeorite impacts and solar wind irradiation
\citep{Yamada:1999,Hapke:2001,Sasaki:2001,Brunetto:2005}.  
The intrinsically dark \cite[p$_v \sim$0.06;][]{Mainzer:2011} 
nature of carbonaceous material 
and the lack of prominent absorption features at visible and near-IR wavelengths
implied space weathering trends would be difficult to identify for 
C-complex asteroids \citep{Hapke:2001,Moroz:1996}.
However, two recent studies indicate 
significant spectral slope variation among this class of asteroids as a function of age 
\citep{Nesvorny:2005,Lazzarin:2006}.  

Principal component analyses of 
asteroid colors in the Sloan Digital Sky Survey (SDSS) show the mean slopes of 
C-complex asteroid families experience a decrease in slopes and become spectrally bluer 
with age at visible wavelengths \citep{Nesvorny:2005}. 
In contrast, a study using visible spectroscopic data from the Small 
Main-Belt Asteroid Spectroscopic Survey 
\cite[SMASSII;][]{Bus:2002}, shows the spectral slopes of the C-complex
asteroid population as a whole increase (redden) with age \citep{Lazzarin:2006}.  
However, when \cite{Lazzarin:2006} limit their analyses to the C-complex asteroid
families they are able to reproduce the slope trends obtained
by \cite{Nesvorny:2005}.  \cite{Lazzarin:2006} suggests the discrepancies 
between the two slope trends
arise from a sampling effect where the average compositions of 
C-complex asteroids are not fully represented in the sampled C-complex families.  

The \cite{Nesvorny:2005} and \cite{Lazzarin:2006}
studies indicate significant evidence of space weathering 
trends among the C-complex asteroids, however the effects of compositional 
variation on these trends is still unclear.  
In the case of S-complex asteroids, \cite{Nesvorny:2005} use 
the Koronis asteroid family and Karin cluster, which formed from 
the breakup of a Koronis family member, to study 
space weathering trends while avoiding the influence of compositional 
variation.  The spectral trends of the 
young Karin \citep[5 Myr;][]{Nesvorny:2002} and old Koronis 
asteroids \citep[2.5 Gyr;][]{Marzari:1995} confirm that the reddening observed among the
S-complex asteroids is indeed a result of space weathering and 
not a product of mineralogical variation.  
The recent discovery of a sub-family of the Themis asteroids 
allows a similar test to be conducted on C-complex asteroids. 
The Themis family \citep{Hirayama:1918} 
has $\sim$4,000 members \citep{Milani:2014} resulting from the catastrophic break up of
a 390-450~km parent body asteroid 2.3 Gyr ago \citep{Marzari:1995}.  
A recent ($<$ 10 Mya) break up of a Themis family 
member (D$\sim$20~km to 65~km in size) resulted in the formation of the 
Beagle family, which contains $\sim$60 asteroids \citep{Nesvorny:2008}.  
These two families are the first C-complex families identified which
originate from the same parent body and provide 
a unique tool for assessing C-complex space weathering trends while 
alleviating the mineralogical variations among C-complex asteroids. 

In this study we use spectroscopic and albedo data to search for space 
weathering trends among the Themis and Beagle asteroids.  We compare these
trends to those of the Veritas asteroid family, which is another young \cite[8.3 Myr;][]{Nesvorny:2003}
C-complex family.  The Veritas asteroids are not related to the Themis and Beagle asteroids
and allow us to assess whether differences in mineralogy affect the spectral trends 
in space weathering studies of C-complex asteroids.   

\section{Observations}
In this paper, we present data for 52 main-belt asteroids 
belonging to the Themis, Beagle and Veritas
asteroid families.  
Observations took place on 3.5 nights during Mar. 03, Oct. 29, Oct. 30, 2013 
and Feb. 21, 2014 at the 8.2-meter Subaru telescope on Maunakea, Hawai'i
(see Table \ref{tab_subaru_observations} for log of observations).
Spectroscopic data covering the 0.47 $< \lambda <$ 0.91 $\mu$m spectral range were 
taken with the Faint Object Camera and Spectrograph (FOCAS).
FOCAS has two 2K$\times$4K CCDs with a total of 8 readout channels, 
each with 512$\times$4176 pixels \citep{Kashikawa:2002}. 
The spectra were recorded in channel 3 of chip 2, which has a read noise of 
3.4 e- and gain of 2.082 e-/ADU.   We used the lowest resolution grating 
(75 gr/mm) and 2x2 binning to obtain the highest signal to noise ratio (SNR) for our 
targets which are dominated by relatively small (D $<$15~km) and faint asteroids. 
The resulting low resolution dispersion of 11.8 \AA/pixel is ideal for detecting 
the shallow and broad absorption features typically seen in C-complex asteroid 
spectra. 
The SY47 order sorting filter was used to prevent higher order contamination.  
Observations were taken through a 1" wide slit, oriented to the parallactic
angle, while tracking at non-sidereal rates.  Due to the lack of 
non-sidereal guiding and to prevent drifting across the slit, 
integration times were limited to 600 seconds.

Family members were selected based on the dynamically derived 
memberships found in \cite{Nesvorny:2012}.  
Previous studies show trends where the frequency of asteroids with
absorption features varies as a function of diameter
\citep{Florczak:1999,Fornasier:2014}, so 
Themis and Veritas targets were limited to members
with similar diameters as Beagle targets (D $\le$ 15~km) to 
avoid size related variations.  Nearby solar analog stars were observed at 
airmasses similar to the asteroids throughout the night and ThAr 
lamp spectra were taken at the beginning 
or end of each observing run for wavelength calibration.

\section{Data Reduction and Analysis}
Data were reduced using the 
Image Reduction and Analysis Facility (IRAF) V2.14 {\tt noao} 
longslit and Subaru {\tt focasred} packages \citep{Tody:1986}.  The reduction procedure included
overscan and bias subtraction, image trimming, flattening, and cosmic ray removal for 
long exposure images. The {\tt apall} package was used to perform background subtraction,
extraction of the


\noindent one-dimensional and associated sigma spectra from the 
two-dimensional images.  The sigma spectrum is produced by measuring the 
sigma at each wavelength in the two-dimensional image. 
Following extraction, data were wavelength calibrated 
using emission lines from the ThAr lamp spectra.

Median combined asteroid spectra were divided by nearby G2V solar analog spectra
to produce reflectance spectra for each asteroid.  The solar analog stars were 
observed close in time to the observations of each asteroid and were chosen to 
have an airmass difference of $<$0.1 to enable extinction correction.  
The reliability of each solar analog star was tested
by producing multiple reflectance spectra of an asteroid 
from multiple stars at similar airmasses.  Only standard stars that produced
consistent spectral shapes and slopes were used in producing reflectance spectra.
Table \ref{tab_subaru_observations} reports which
solar analog stars were used to produce the reflectance spectra for each asteroid.  
Residual features due to incomplete removal of sky lines 
were removed using a median filter to search for erroneous spikes 
in the final reflectance spectra.

We measured slopes using the equation for the normalized reflectivity 
gradient S$'$ (reported in units of  \%$/$1000 \AA), as defined by \cite{Jewitt:1986}:
\begin{equation}\label{slope}
S' (\lambda_1,\lambda_2)=\left( \frac{dS / d\lambda}{S_{0.55}} \right)
\end{equation}
where dS$/$d$\lambda$ is the slope of the reflectivity measured within 
the wavelength region between $\lambda_1$ and $\lambda_2$. 
Reflectance spectra were normalized to 0.55~$\mu$m and 
spectral slopes were measured using a weighted linear least-squares fit to 
data between 0.49 $< \lambda < 0.91~\mu$m.  
The fits were weighted using the IRAF generated sigma spectrum for each asteroid.  
Data shorter than 0.49~$\mu$m were not included in the fit due to 
potential contamination from the UV absorption below $\sim$0.5~$\mu$m
created by an Fe$^{2+}$ intervalence charge transfer attributed to 
phyllosilicates \citep{Vilas:1994,Hiroi:1996}.  
Asteroid slopes and 1-sigma uncertainties are shown 
in Table \ref{tab_subaru_data}.  


At large phase angles, asteroids are known to experience an increase in slopes (reddening) 
\citep{Millis:1976,Bowell:1979,Lumme:1981}, however, \cite{Luu:1990} 
find that adopting phase coefficients from 
other C-complex asteroids does little to improve slope measurements
for observations at small phase angles ($\lesssim40$). 
In addition, we do not have phase reddening coefficients for most of our targets, but 
our observations span a small range of phase angles 
($\lesssim$20$^{\circ}$), thus no phase 
reddening correction was applied to our reflectance spectra.

\begin{figure*}[h]
\begin{center}
\includegraphics[width=0.75\linewidth,angle=90]{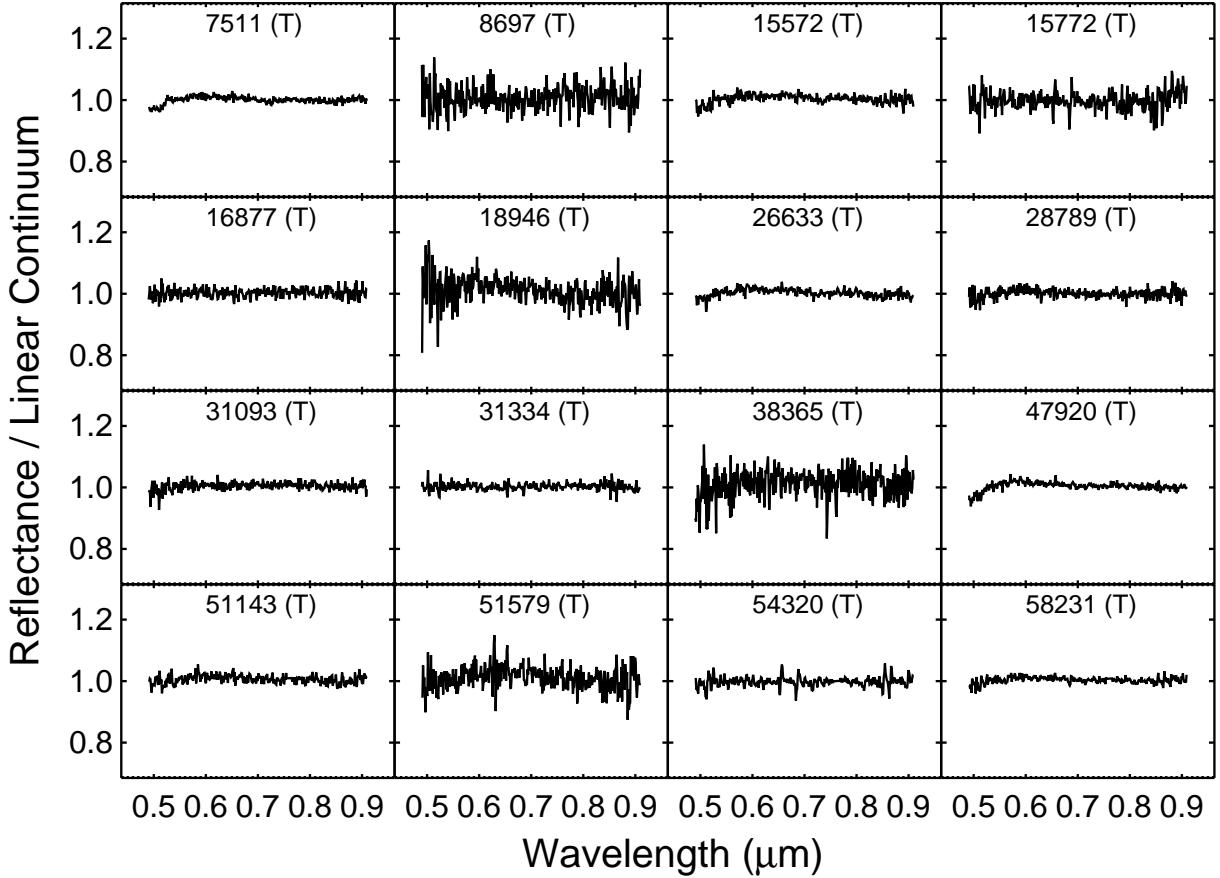} 
\caption{\label{subaru_all_cr1} Continuum removed spectra of individual Themis (T) asteroids,
where labels report the asteroid ID. }
\end{center}
\end{figure*}

\begin{figure*}[h]
\begin{center}
\includegraphics[width=0.75\linewidth,angle=90]{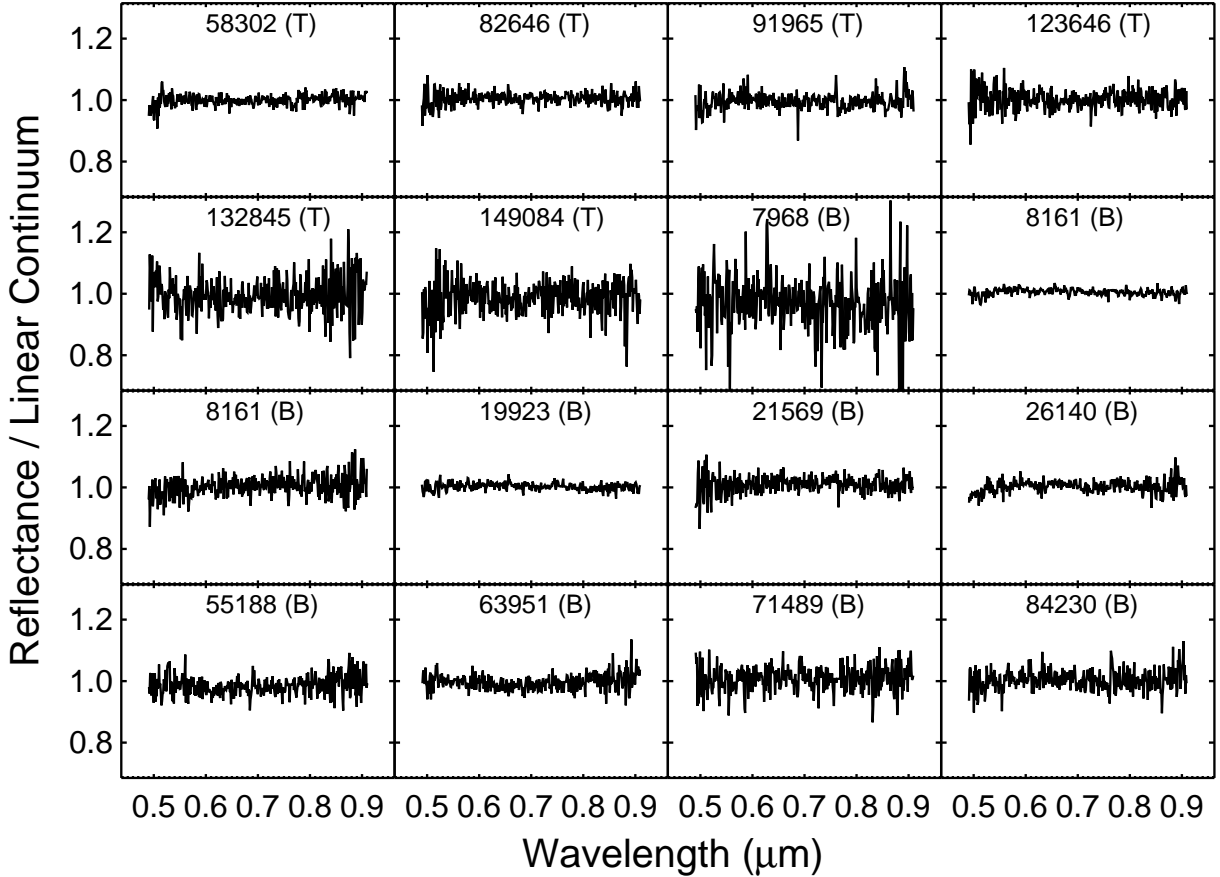} 
\caption{\label{subaru_all_cr2} Continuum removed spectra of individuals asteroids.
Labels show the asteroid ID as well as family membership, where (T) and (B) are Themis, 
Beagle family members, respectively.}
\end{center}
\end{figure*}

\begin{figure*}[h]
\begin{center}
\includegraphics[width=0.75\linewidth,angle=90]{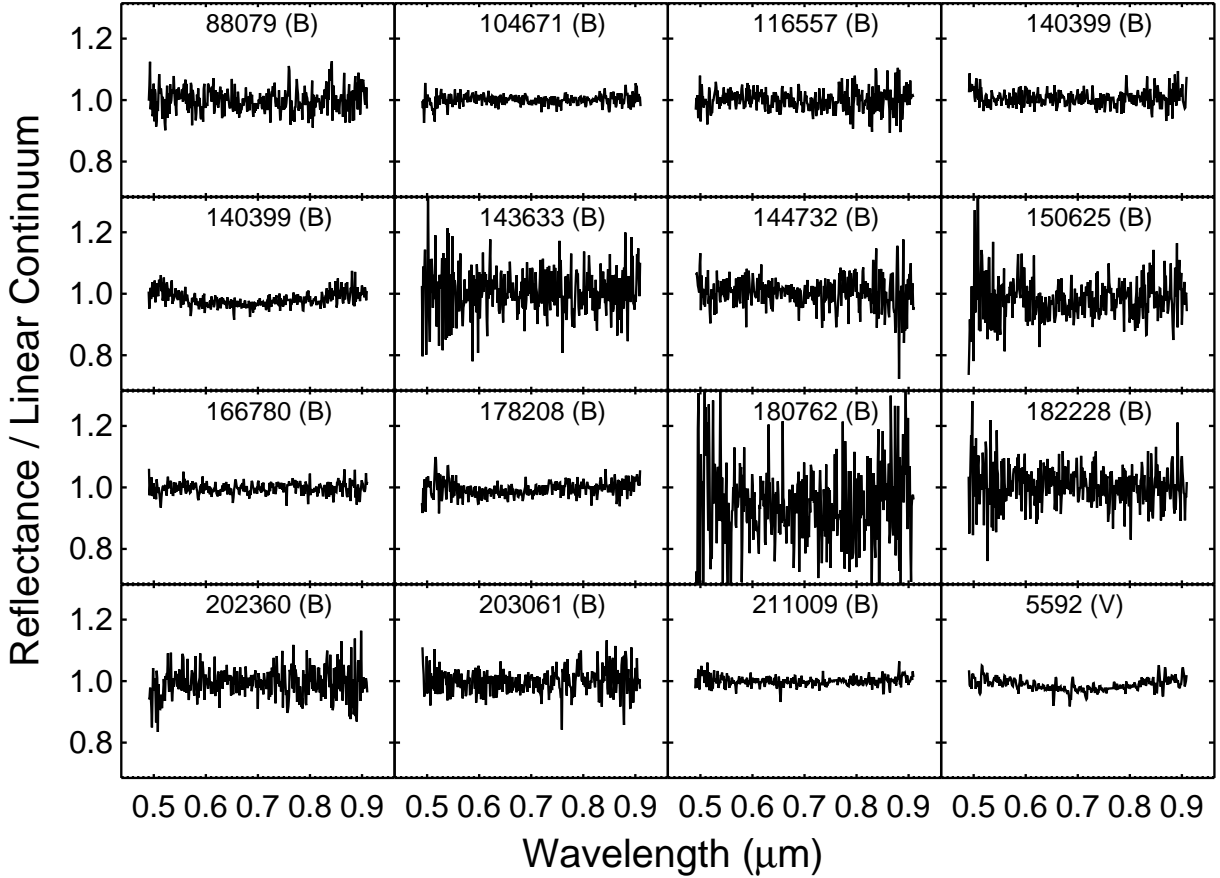} 
\caption{\label{subaru_all_cr3} Continuum removed spectra of individuals asteroids.
Labels show the asteroid ID as well as family membership, where 
(B) and (V) are Beagle and Veritas family members, respectively. }
\end{center}
\end{figure*}

\begin{figure*}[h]
\begin{center}
\includegraphics[width=0.75\linewidth,angle=90]{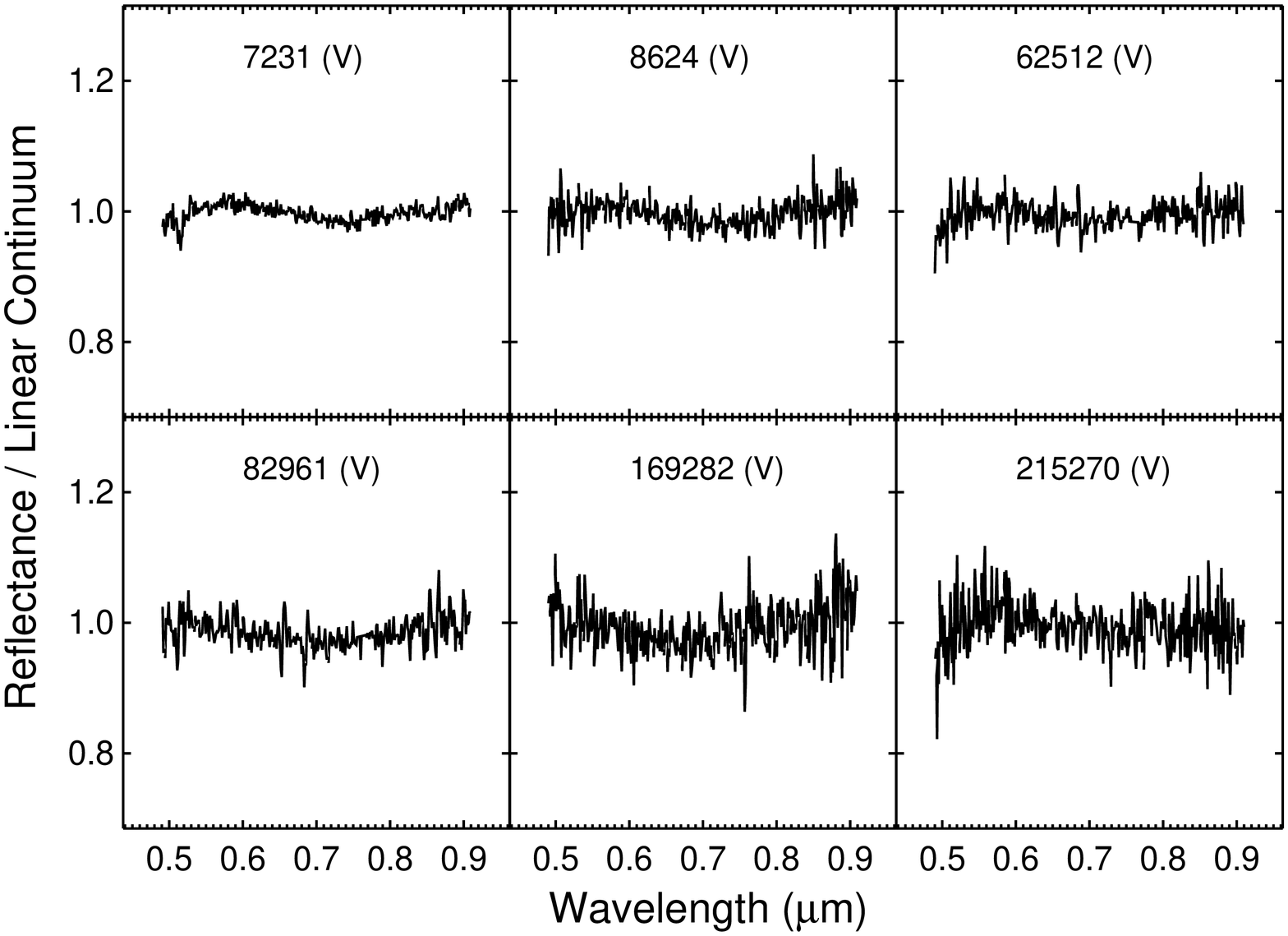} 
\caption{\label{subaru_all_cr4} Continuum removed spectra of individual Veritas (V) asteroids,
where labels report the asteroid ID. }
\end{center}
\end{figure*}

One of the few features found in C-type spectra is a 
shallow ($<$ 5\%) 0.7 $\mu$m absorption band created by phyllosilicates \citep{Vilas:1989}.  
To search for this feature we perform continuum removal and fit 
the 0.55-0.85 $\mu$m feature region with a 2nd or 3rd order
polynomial.  The continuum removed spectra were created by dividing the full spectrum 
by a linear fit to the 0.7~$\mu$m continuum shoulders (0.52-0.54~$\mu$m and 
0.86-0.88~$\mu$m).  The continuum removed spectra 
are shown in Figures \ref{subaru_all_cr1}-\ref{subaru_all_cr4}.  
The band depth and center were determined using the reflectance value at the wavelength 
corresponding to the minimum of the polynomial fit.  The band depth uncertainty was 
derived using the uncertainty in the sigma spectrum corresponding to the position of the band 
center and only features with band depths greater than the associated uncertainties were
flagged as a detection.  Band depths for objects where features were detected and 
sensitivity limits computed from the SNR of each spectrum are reported in Table 
\ref{tab_subaru_data}.

In addition to our Subaru data, we use
visible albedos derived by \cite{Masiero:2011} from the Wide-field Infrared Survey 
Explorer (WISE) to characterize the albedo distributions of Themis, Beagle
and Veritas asteroids.  We used 2025, 565, and 16 Themis, Veritas and Beagle members,
respectively, in our analyses but divided the data set into two size 
populations (cutoff = 15~km) for consistency with our Subaru data.  
The mean values for slope and albedo, the standard error of the mean (SEM) and the standard deviations 
for each family and size range are reported in Table \ref{tab_family_summary}.  We use the 
SEM to quantify the variation in our estimate of the mean, and the 
standard deviation to quantify the dispersion in our slope and albedo distributions. 

\section{Results}
Members of the Themis and Beagle families show a wide range of 
blue to red-sloped spectra (Table \ref{tab_subaru_data}).  However, 
$\sim$90\% (21 out of 23) Beagle asteroids have blue-sloped spectra, 
whereas 60\% (13 out of 22) of the Themis asteroids are blue-sloped. 
These data suggest the Beagle family is dominated by B-type asteroids, which 
are characterized by negatively sloped (blue) spectra \citep{Tholen:1984,Bus:2002a}. 
The Themis asteroids show a range of C, B, F and G spectral types \citep{Tholen:1984}, 
consistent with the observations of other Themis members \citep{Florczak:1999}.

\subsection{Spectral Slopes}
\begin{figure*}[h!]
\begin{center}$
\begin{array}{cc}
\includegraphics[width=0.6\linewidth,angle=90]{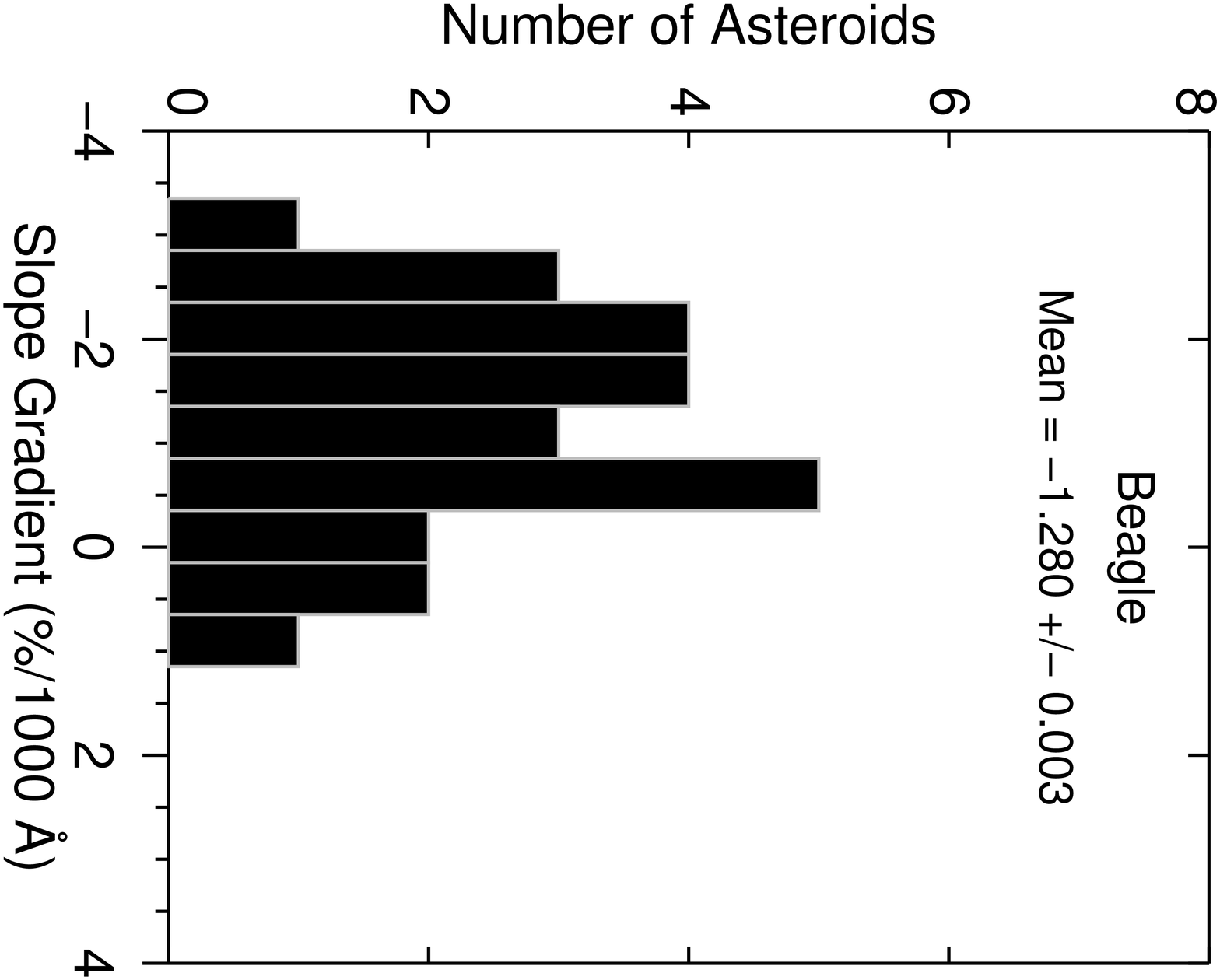} 
\includegraphics[width=0.6\linewidth,angle=90]{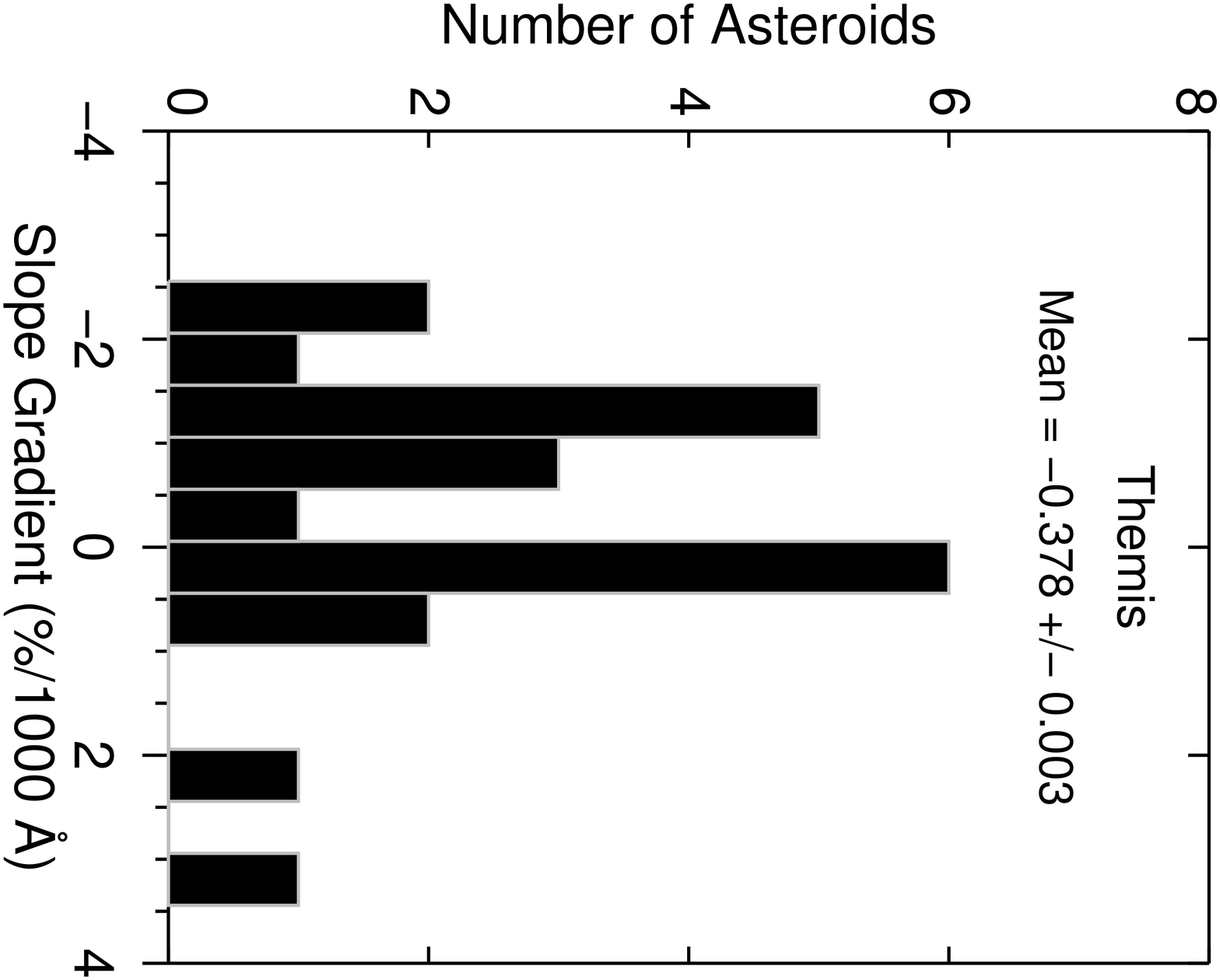} 
\end{array}$
\end{center}
\caption{\label{Slopes_Histogram}  Spectral slope gradients (as defined by \cite{Jewitt:1986}) 
for Themis and Beagle members.  The weighted mean and 
standard error of the mean are reported.
}
\end{figure*}

The slope distributions for the Themis and Beagle families are shown 
in Figure \ref{Slopes_Histogram}.  
Given the larger fraction of blue sloped spectra, 
the weighted mean for the Beagle asteroids ($-1.280$ $\pm$ 0.003\% per 1000~\AA) 
is significantly bluer than that of the Themis asteroids  ($-0.378$ $\pm$ 0.003\% per 1000~\AA).
The difference in slopes between Themis and Beagle imply these asteroids redden 
by $\sim$0.9\% per 1000~\AA~over a timespan of $\sim$2.3 billion years. 
To test the significance of the difference in the slope distributions of the two families, we applied 
the Student's T and Kolmogorov-Smirnov (KS) tests to the Themis and Beagle 
data sets.  We find the difference in means is significant at the 2-sigma level.

The Themis asteroid 18946 is extremely red (3.037 $\pm$ 0.005\% per 1000~\AA),
and is located near the outer region of the proper element distributions for 
the Themis family.  Due to the possibility of 18946 being an interloper, we
exclude 18946 from our analyses, and compute a new mean 
slope of $-0.505$ $\pm$ 0.003\% per 1000~\AA~for the Themis asteroids.
The reduced mean equates to a slope reddening rate of
$\sim$0.8\% per 1000~\AA~in $\sim$2.3 billion years. 
After re-applying the statistical tests, the difference in means between the Themis and 
Beagle populations is still significant at the 2-sigma level although there 
is a slight drop in significance.   

The Beagle asteroids 140399 and 8161 were each observed on two nights and 
have significantly varied spectral slopes (Table \ref{tab_subaru_data}).  
Both spectra of 140399 show the asteroid 
is moderately blue, however spectra of 8161 are both blue and red sloped.  Both 
spectra were taken during photometric conditions and produced using the same 
solar analog star, thus the 
large slope variation in asteroid 8161 may imply color differences across the surface of 
this asteroid. However, follow-up observations are needed to assess whether 
these trends can be reproduced.

\subsection{Hydration Features}
\cite{Florczak:1999} and \cite{Fornasier:2014} suggest that the 0.7 $\mu$m feature is 
less frequent among asteroids with decreasing diameters.  
Data from the SMASS II survey show a high fraction of 
hydrated members in the young \cite[8.3 Myr;]{Nesvorny:2003} Veritas asteroid family \citep{Bus:2002}, 
so we took spectra of 5 randomly selected small (D $<$ 10 km) and 2 moderately sized (D$\sim$20 km)
Veritas members to test if the 0.7 $\mu$m feature was observable on small diameter objects. 
We found 4 out of 7 (57\%) asteroids less than 25~km in diameter show the 0.7~$\mu$m feature.
Although asteroid 169282 does not pass our detection criteria (the depth of the band is less than
the SNR limits), visual inspection of the spectrum (Fig. \ref{subaru_all_cr4}) does 
suggest the feature is present.  Including asteroid 169282, 
3 out of 5 members ($\sim$60\%) with D $< $15~km show the 
0.7~$\mu$m feature, suggesting we should expect to detect the feature 
on small Themis and Beagle asteroids if phyllosilicates were present.  

However, only one Beagle (140399) and no 
Themis asteroids show the feature.  
It is important to note that asteroid 140399 was imaged on two separate observing 
runs and only one of the spectra (Fig. \ref{subaru_all_cr2}) shows the absorption.
Despite the lack of 0.7~$\mu$m features in the Themis members,
a small number of objects (4 of 22, or 18\%) show evidence of the  Fe$^{2+}$ UV absorption band
also attributed to phyllosilicates \citep{Vilas:1994}.  A similarly small fraction 
(3 of 23, or 13\%) of the Beagle asteroids also show this UV feature. 
Due to the cutoff in the spectral coverage below
0.47~$\mu$m, we did not attempt to fit the Fe$^{2+}$ UV feature, but its presence
was assessed through visual inspection.  Table \ref{tab_subaru_data} shows
which asteroids were flagged as showing the UV feature.

\subsection{Albedo Variations}
\begin{figure*}[t!]
\begin{center}
\includegraphics[width=0.9\linewidth]{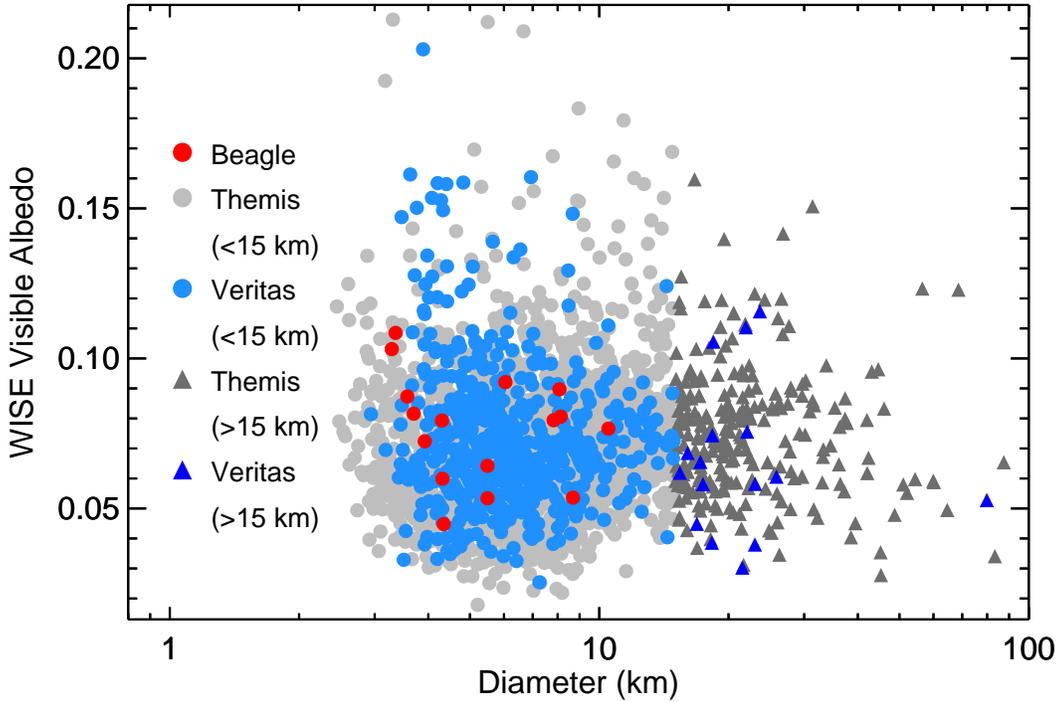} 
\caption{\label{wise_tbv} WISE albedo distributions for Beagle, Themis and 
Veritas asteroid families from \cite{Masiero:2013}.  }
\end{center}
\end{figure*}

The WISE survey sensitivity is strongly dependent on the 
distance and diameters of surveyed asteroids.  As 
Beagle and Themis members have nearly identical orbital elements, the detection 
sensitivity in our analyses is limited to asteroid diameters.  Therefore,  
due to the small size range of the Beagle asteroids, we limit the comparison between
the three families to objects with diameters less than 15~km.  
Figure \ref{wise_tbv} shows the WISE albedo 
distributions as function of diameter for both the small (D $\le$ 15~km) and large 
(D $>$ 15~km) asteroid populations in each family.  The population of small
Themis asteroids have a lower average albedo (p$_{v}$ = 0.068$\pm$0.001) than 
the Beagle asteroids (p$_{v}$ = 0.079$\pm$0.005).  
Unexpectedly, the small diameter Veritas asteroids, which are similar in age to the Beagle family,
have albedos (p$_{v}$ = 0.069$\pm$0.001) similar to the older Themis asteroids within the 
same size range.  In addition, the larger Themis asteroids (D $>$ 15~km) 
are significantly brighter (p$_{v}$ = 0.075$\pm$0.001) than the smaller 
members.  The Veritas asteroids also show some evidence
in albedo variations between small and large members, however the difference
is not statistically significant.

\begin{figure*}[ht!]
\begin{center}
\includegraphics[width=0.8\linewidth]{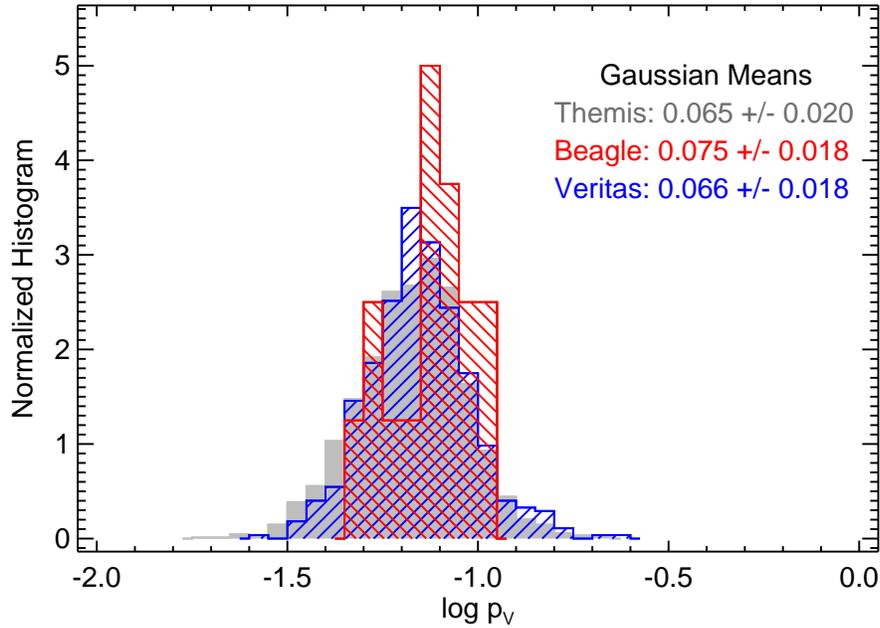} 
\caption{\label{wise_gaussian} Log plot of WISE albedo (p$_{v}$) distributions for Themis, 
Beagles and Veritas asteroids.  Each histogram has
been normalized to unit area for comparison.  
Gaussian mean and widths are reported for each family. }
\end{center}
\end{figure*}

For comparison between our results and \cite{Masiero:2013}, 
we fit gaussian profiles to the log distributions of the three families.  The distributions in log space 
are shown in Figure \ref{wise_gaussian}. 
We find that our gaussian means are consistent with the values derived 
by \cite{Masiero:2013}.  When comparing the gaussian medians with 
the observed sample distribution, we find the gaussian values are 
slightly larger than the means of the observed population.  The difference in means imply 
the observed population does not follow a gaussian distribution.  
The apparent non-gaussian distribution may originate from 
an observational bias introduced by the dependence of WISE derived 
albedos upon previously published absolute magnitudes \citep{Masiero:2013}.   
However, since the Themis, Beagle and Veritas families are located 
at similar heliocentric distances, the observational biases are similar 
for each family, allowing comparison of the 
observed albedo distributions for these families.

\section{Discussion}
Spectral characteristics such as slopes and absorption features 
are one of the few tools available for assessing asteroid 
compositions remotely.  The abundance of the absorber, 
size distribution of the regolith grains \citep{Clark:1999}, and 
space weathering processes \citep{Conel:1970,Gold:1970,Hapke:1970}
all greatly influence the features of asteroid spectra.
Although these processes and characteristics are difficult to 
disentangle when analyzing spectroscopic data, 
the sensitivity of spectroscopy to these processes 
allows it to be a powerful remote sensing tool.

\subsection{C-Complex Space Weathering Trends}
As seen in Figure \ref{Slopes_Histogram}, the older Themis asteroids 
have distinctly redder slopes than the younger 
Beagle asteroids.  This difference implies the spectral slopes 
of C-complex asteroids become redder over time, which is consistent 
with the trends observed by \cite{Lazzarin:2006}.  Although the slope
evolution of C-complex asteroids appears consistent with silicate-rich 
asteroids \citep{Nesvorny:2005,Lazzarin:2006}, the 
albedo trends are not.  \cite{Masiero:2011} find the younger Karin 
family has a lower albedo than the older Koronis family.  In contrast,
our data suggest the younger Beagle asteroids have a higher albedo 
than the older Themis asteroids.  Future work is necessary to assess
whether the albedo trends in Karin and Koronis are representative of the broad S-complex 
weathering trends.   

The relatively young ages of the Veritas and Beagle asteroids 
suggest they have minimally evolved regoliths, yet 
the Veritas asteroids show a much higher fraction (57\%) of hydrated members
than the Beagle asteroids (13\%).  Therefore, the low fraction of features in the 
Beagle asteroids likely reflects an intrinsically low phyllosilicate content.
However, considering the higher fraction (35-60\%) 
of large Themis members with phyllosilicate features
\citep{Florczak:1999}, we postulate the diagnostic features have been removed due to 
regolith evolution and space weathering processes.  We describe this hypothesis 
further in $\S$ \ref{sec:size trends}.  In addition to the difference in phyllosilicate features, 
the Veritas family albedos show that they are darker than the Beagle 
asteroids, but comparable to the older Themis asteroids.   
Thus, these data imply a significant difference in the composition 
between the Veritas and Themis/Beagle families and that the 
trend observed by \cite{Nesvorny:2005} is likely a product
of mineralogical variation rather than space weathering processes.  

Classical space weathering models attribute the reduced band depths, darkening and reddening of 
atmosphereless bodies to the optical effects of SMFe particles \citep{Hapke:2001}. 
Although C-complex asteroids appear distinct in their mineralogy 
when compared to lunar soils and S-complex 
asteroids \citep{Bus:2002b}, many of the visible wavelength absorption features in C-complex asteroids
are attributed to oxidized Fe transitions \cite[][and references within]{Rivkin:2002}.
Thus, the availability of Fe and our observed spectral trends suggest space 
weathering processes on C-complex asteroids results in the production of SMFe particles. 

Using the age difference between the Themis and Beagle 
families, we estimate a slope increase of $\sim$0.8\% per 1000~\AA~
or $\sim$0.08 $\mu$m$^{-1}$ in 2.3 Gyr for C-complex asteroids.  In comparison, 
\cite{Lazzarin:2006} estimates a weathering rate of 
8.8~$\pm$~4.6 $\times$ 10$^{-5}$ $\mu$m$^{-1}$ AU$^2$ for the 
C-complex asteroid population, which in 2.3 Gyr
equates to $\sim$0.02$\mu$m$^{-1}$ for objects with circular 
orbits located near the Themis asteroids ($\sim$3.1AU).  For S-complex 
asteroids, \cite{Lazzarin:2006} estimates a rate of 24.9~$\pm$~4.6 
$\times$ 10$^{-5}$ $\mu$m$^{-1}$ AU$^2$, corresponding to $\sim$0.06 $\mu$m$^{-1}$ in 
2.3 Gyr for objects at 3.1 AU.  Although no estimate was derived for 
C-complex asteroids, \cite{Nesvorny:2005} derive a weathering rate for S-complex asteroids 
of 0.01 $\mu$m$^{-1}$ $\times$ $\log_{10}t$ ($t$ given in Myr), which
equates to $\sim$0.03 $\mu$m$^{-1}$ in 2.3 billion years.  In contrast, 
\cite{Vernazza:2009} suggests space weathering occurs 
very rapidly within the main asteroid belt, finding that S-complex asteroid 
slopes redden by as much as $\sim$0.4 $\mu$m$^{-1}$ in several million 
years for newly formed families.  A direct comparison between each of these studies 
is difficult due to the inherent difference in each data set, 
likely causing the significant difference between each of the estimates.  However,
our derived rates support the estimates of \cite{Lazzarin:2006} 
and \cite{Nesvorny:2005}, which suggest that space weathering is a 
long term process for both S and C-complex main belt asteroids.

\subsection{Size Related Spectral Trends}
\label{sec:size trends}
Several groups note trends of a decreased frequency in 
phyllosilicate absorption features with decreasing size of asteroids
\citep{Jones:1990,Vilas:1996,Howell:2001,Fornasier:2014}. 
The same trend has been noted in the Themis asteroids in particular, 
with $\sim$60\% of asteroids greater than 50~km presenting phyllosilicate features 
in contrast to 35\% for objects less than 50~km.  
\citep{Florczak:1999}.  Additionally, our data show that the 
fraction reduces to 18\% for Themis asteroids with diameters less than 15~km.  
Furthermore, these size related trends also extend to Themis family albedos, 
where objects with less than 15~km are 
darker than larger asteroids.

Due to their lower escape velocities and the inability to retain fine particles during collisions, 
small asteroids are thought to develop 
relatively thin regoliths dominated by coarse grains \citep{Matson:1977,Chapman:1978}.
Indeed, spacecraft observations of small asteroids such as Eros and Itokawa 
reveal surfaces dominated by gravel and meter sized boulders with fines constricted to 
low potential zones \citep{Chapman:2002a,Miyamoto:2007}.  
These rocky surfaces are hypothesized to be
more resistant to gardening processes \citep{Chapman:2002,Willman:2010}, thus 
space weathering effects (decreased albedo and band depths) 
are likely to be more prevalent on small objects with time. 
Conversely, impacts upon large asteroids with thicker regoliths and finer particles
may be more effective in counteracting space weathering through regolith gardening 
\citep{Willman:2010}, thereby preserving spectral features for longer. 
Furthermore, laboratory data show that increasing grain size 
results in a decreased reflectance or albedo and a reduction in absorption band depths 
\citep{Fischer:1994,Clark:1999,Cloutis:2011a,Cloutis:2011}. 
We propose the size related absorption feature and albedo trends in the
Themis asteroids originate from the combined effects of coarse grained regoliths and 
optical maturation times, which result in a pronounced reduction of absorption 
features and decreased albedos for smaller asteroids.

In contrast to the spectral trends seen in the Themis asteroids, 
\cite{Binzel:1996} found that the small near-Earth asteroids (NEAs) 
exhibit deeper absorption bands when compared to large NEAs.  \cite{Chapman:2004}
suggests these trends originate from the rapid erosion and resetting of regoliths
on smaller asteroids.  However, impact velocities are lower 
in the main belt than the near Earth region \citep{Bottke:2002}, thus 
impact processes will different between the two populations and 
may explain the difference in spectral trends with size.
Further observations are necessary to assess 
whether the albedo and absorption feature trends seen in the Themis 
asteroids are found among
other main belt asteroid families.  

\subsection{Aqueous Alteration in the Themis Parent Body}
Several main-belt comets (MBC) \citep{Hsieh:2006a}
have been dynamically linked to the Themis family 
\citep{Hsieh:2009,Novakovic:2012}.  The source of 
sublimation in these MBCs is attributed to 
water ice buried beneath a layer of regolith several 
meters thick \citep{Schorghofer:2008,Prialnik:2009}.
Thermal evolution models suggest the Themis/Beagle 
parent body was likely differentiated, with aqueous alteration 
taking place predominantly in the core, thus allowing the preservation of 
an icy shell near the surface \citep{Castillo-Rogez:2010,Grimm:1989,Cohen:2000}. 

Due to the young age of the Beagle family and
the small fraction of hydrated Beagle members,
we believe that the low frequency of features 
represents an intrinsically low phyllosilicate content
in the Beagle parent body.  The 
detection of phyllosilicates in the Beagle asteroids has important
implications for geophysical models of the Themis/Beagle parent
body, considering the MBC 133P/Elst-Pizarro is a possible 
member of the Beagle family \citep{Nesvorny:2008}.  If indeed a member,
then the presence of 133P and its volatiles
implies the Beagle parent body originated from an 
ice-rich layer in Themis/Beagle parent body.  However,
the small but notable fraction of members with the UV and 0.7~$\mu$m 
features suggest the Beagle parent body 
was not homogeneous, but rather a mixture of ice and 
aqueously altered material.  

We propose two possibilities for heterogeneity of the Beagle parent body. 
The first is that heterogeneity may represent non-uniform alteration of the 
Themis/Beagle parent body.  Conversely, \cite{Rivkin:2010} and \cite{Campins:2010}
have potentially detected water ice features on the largest member of the family, 24 Themis. 
\cite{Castillo-Rogez:2010} suggest 24 Themis is the remnant silicate rich
core onto which components of its icy crust may have reaccreted 
following the collisional disruption of the parent body.  It is possible 
the Beagle parent body may be an
object which reaccreted non-native components, resulting in  
the present heterogeneity of the Beagle asteroids.  Exploring 
the viability of these two scenarios is beyond the scope of this paper, however
coupled geophysical and dynamical modeling of the two families may provide 
valuable insights into the evolution of this body. 

\section{Conclusions}
Through comparison of visible spectra and albedo data of the Themis, Beagle 
and Veritas asteroids, we find the following: 

\begin{itemize}

\item Space weathering results in an increase of spectral slopes and a 
decrease in albedo among C-complex asteroids. 

\item The trends in decreased phyllosilicate features and albedos with 
decreasing diameter likely result from variations in regolith properties
as a function of age and diameter.  

\item The notable fraction of Beagle members with phyllosilicate features
indicate the Beagle parent was a heterogeneous mixture of ice and 
aqueously alterated minerals. 

\item The apparent mineralogical differences
between the Veritas family and the Themis and Beagle families highlight
the importance of accounting for mineralogy when interpreting space weathering trends 
across the broad population of C-complex asteroids.

\end{itemize}
 
\acknowledgments
We would like to say mahalo nui loa to 
Jan Kleyna, Takashi Hattori and Bin Yang for their help and support of this work. 
This work is based upon data collected at Subaru Telescope, which is operated by 
the National Astronomical Observatory of Japan. 
Image processing in this paper has been performed using the IRAF software.  IRAF 
is distributed by the National Optical Astronomy Observatories, which is 
operated by the Association of Universities for Research in Astronomy, Inc. 
under cooperative agreement with the National Science Foundation.  
Support for this work was provided by the National Aeronautics
and Space Administration through the NASA Astrobiology Institute under
Cooperative Agreement No. NNA04CC08A issued through the Office of Space
Science, by NASA Grant No. NNX07A044G. 

{\bibliography{SW_C-types_ArXiv.bib}}
\begin{deluxetable}{lcrc cccrccccc}
\tablecolumns{13}
\tabletypesize{\scriptsize}
\tablewidth{0pt}
\tablecaption{\large \bf Observing Geometry and Conditions \label{tab_subaru_observations}}
\tablehead{
\colhead{\bf Asteroid}   &
\colhead{\bf Vmag}& 
\colhead{\bf a}& 
\colhead{\bf e} &
\colhead{\bf i} &
\colhead{\bf r$^\dag$} &
\colhead{\bf $\Delta^\ddag$} &
\colhead{\bf $\alpha^\S$}  &
\colhead{\bf Obs} &
\colhead{\bf No. of} & 
\colhead{\bf Airmass} &
\colhead{\bf Solar}  &
\colhead{\bf Airmass}  \\
\colhead{\bf ID} &
\colhead{}&
\colhead{\bf (AU)}&
\colhead{} &
\colhead{\bf (deg)} &
\colhead{\bf (AU)}&
\colhead{\bf (AU)}&
\colhead{\bf (deg)}&
\colhead{\bf Run}&
\colhead{\bf Spectra}&
\colhead{\bf (Object)} &
\colhead{\bf Analog}& 
\colhead{\bf (Stand)} 
}
\startdata
\hline
{\it Themis } &&& &&& &&& &&& \\
\hline
7511  & 17.5  & 3.205  & 0.157  & 1.277  & 2.694  & 2.023   &  6.739 &  3 &  2 &  1.23 &  HD19061 &  1.29  \\     
8697  & 18.7  & 3.106  & 0.162  & 0.978  & 3.606  & 2.637   &  3.641 &  4 &  2 &  1.04 &  HD73708 &  1.03  \\     
15572  & 18.1  & 3.209  & 0.168  & 2.109  & 3.030  & 2.131  &  9.525 &  3 &  2 &  1.30 &  HD7983 &  1.28  \\      
15772  & 19.6  & 3.217  & 0.143  & 2.694  & 3.530  & 3.265  &  16.127 &  4 &  2 &  1.06 &  HD284013 &  1.02  \\   
16877  & 17.6  & 3.055  & 0.122  & 1.307  & 2.726  & 1.775  &  7.075 &  4&  2 &  1.04 &  HD73708 &  1.03  \\      
18946  & 19.3  & 3.046  & 0.178  & 0.522  & 3.181  & 2.281  &  8.836 &  1 &  1 &  1.46 &  HD73708 &  1.39  \\     
26633  & 18.2  & 3.220  & 0.176  & 2.544  & 2.702  & 1.852  &  12.816 &  3 &  2 &  1.27 &  HD7983 &  1.28  \\     
28789  & 18.9  & 3.132  & 0.172  & 0.513  & 2.857  & 1.964  &  10.830 &  3 &  2 &  1.03 &  HD19061 & 1.11  \\     
31093  & 17.7  & 3.135  & 0.162  & 3.119  & 3.144  & 2.256  &  9.348 &  4 &  2 &  1.44 &  G104-335 &  1.45  \\    
31334  & 17.7  & 3.176  & 0.132  & 1.436  & 2.807  & 1.865  &  7.536 &  4 &  2 &  1.03 &  HD73708 &  1.03  \\     
38365  & 18.9  & 3.198  & 0.194  & 2.558  & 3.275  & 2.308  &  4.506 &  1 &  1 &  1.49 &  BD+30 2047 &  1.34  \\  
47920  & 18.0  & 3.126  & 0.145  & 2.967  & 3.186  & 2.206  &  3.570 &  3 &  2 &  1.03 &  HD19061 &  1.02  \\     
51143  & 18.4  & 3.146  & 0.167  & 1.178  & 3.139  & 2.213  &  8.052 &  3 &  2 &  1.03 &  HD19061 &  1.11  \\     
51579  & 19.4  & 3.166  & 0.196  & 2.688  & 2.610  & 2.416  &  22.362 &  3 &  2 &  1.35 &  HD197081 &  1.26  \\   
54320  & 17.8  & 3.116  & 0.186  & 0.635  & 2.543  & 1.559  &  3.519 &  3 &  2 &  1.13 &  HD19061 &  1.02  \\     
58231  & 17.6  & 3.222  & 0.156  & 2.033  & 2.735  & 1.744  &  1.067 &  3 &  2 &  1.16 &  HD19061 &  1.11  \\     
58302  & 18.3  & 3.097  & 0.156  & 1.607  & 2.710  & 1.808  &  10.675 &  1 &  1 &  1.27 &  HD73708 &  1.39  \\    
82646  & 18.1  & 3.061  & 0.163  & 3.381  & 2.692  &  1.740 &  7.111 &  4 &  2 &  1.11 &  HD73708 &  1.03  \\     
91965  & 18.7  & 3.244  & 0.113  & 1.381  & 2.977  & 2.091  &  10.189 &  1 &  1 &  1.41 &  HD73708 &  1.39  \\    
123646  & 18.5  & 3.227  & 0.117  & 2.436  & 2.901  & 1.911 &  1.171 &  1 &  1 &  1.66 &  HD95868 &  1.55  \\     
132845  & 20.1  & 3.122  & 0.175  & 1.516  & 2.580  & 1.993 &  20.330 &  3 &  2 &  1.64 &  HD199011 &  1.66  \\   
149084  & 19.7  & 3.180  & 0.166  & 2.603  & 2.676  & 1.840 &  13.769 &  1 &  1 &  1.35 &  HD73708 &  1.39  \\   
\hline
{\it Beagle  } &&& &&& &&& &&&\\
\hline
7968  & 20.5  & 3.157  & 0.163  & 1.386  & 2.856  & 2.233      &  17.417 &  3 &  2 &  1.23 &  HD220764 &  1.27  \\  
8161  & 17.0  & 3.167  & 0.168  & 2.544  & 2.778  & 1.787      &  1.565 &  2 &  2 &  1.25 &  HD19061 &  1.20  \\    
 	& 19.2  & --  & -- & -- & 2.944  &  3.113                     &  18.518 &  4 &  2 &  1.21 &  HD19061 &  1.11  \\   
19923  & 18.5  & 3.161  & 0.111  & 1.698  & 2.839  & 2.068     &  14.693 &  4 &  2 &  1.15 &  G104-335 &  1.13  \\  
21569  & 19.2  & 3.164  & 0.116  & 0.676  & 2.849  & 2.578     &  20.266 &  1 &  1 &  1.37 &  G104-335 &  1.33 \\   
26140  & 19.2  & 3.152  & 0.137  & 2.545  & 3.419  & 2.563     &  9.435 &  3 &  2 &  1.29 &  HD7983 &  1.28  \\  
55188  & 18.6  & 3.160  & 0.183  & 2.453  & 2.807  & 2.407     &  20.152 &  2 &  2 &  1.00 &  HD73708 &  1.04  \\   
63951  & 20.2  & 3.154  & 0.180  & 0.454  & 2.693  & 2.456     &  21.632 &  3 &  2 &  1.12 &  HD73708 &  1.20  \\   
71489  & 19.7  & 3.150  & 0.130  & 0.781  & 3.420  & 2.966     &  15.844 &  3 &  2 &  1.27 &  HD603 &  1.28  \\     
84230  & 20.0  & 3.157  & 0.167  & 0.925  & 2.911  & 2.194     &  15.416 &  3 &  2 &  1.13 &  HD220764 &  1.28  \\  
88079  & 19.2  & 3.151  & 0.139  & 0.112  & 2.761 & 1.870      &  10.939 &  1 &  1 &  1.06 &  HD73708 &  1.09  \\   
104671  & 19.5  & 3.162  & 0.162  & 0.852  & 3.061  & 2.068    &  0.397 &  2 &  2 &  1.04 &  HD19061 &  1.02  \\    
116557  & 20.2  & 3.161  & 0.165  & 1.203  & 2.934  & 2.177    &  14.392 &  3 &  2 &  1.35 &  HD220764 &  1.27  \\  
140399  & 20.0  & 3.146  & 0.174  & 1.701  & 2.705  & 2.719    &  21.053 &  3 &  2 &  1.22 &  HD73708 &  1.20  \\   
 	       & 18.7  &  -- & --  & --        		 & 2.870  & 1.905    &  5.227 &  4 &  2 &  1.60 &  HD95868 &  1.28  \\    
143633  & 20.1  & 3.160  & 0.109  & 0.755  & 2.912  & 2.004    &  9.276 &  4 &  2 &  1.13 &  G104-335 &  1.13  \\   
144732  & 20.6  & 3.152  & 0.134  & 0.435  & 3.011  & 2.305    &  15.249 &  2 &  3 &  1.45 &  HD284013 &  1.29  \\  
150625  & 20.3  & 3.150  & 0.181  & 1.353  & 3.179  & 2.188    &  0.077 &  1 &  1 &  1.43 &  HD95868 &  1.39  \\    
166780  & 19.1  & 3.159  & 0.170  & 1.343  & 2.695  & 1.702    &  0.866 &  2 &  2 &  1.09 &  HD19061 &  1.02  \\    
178208  & 19.8  & 3.172  & 0.113  & 2.565  & 3.056  & 2.070    &  1.835&  4 &  2 &  1.10 &  HD95868 &  1.05  \\     
180762  & 20.9  & 3.147  & 0.175  & 2.183  & 3.659  & 2.697    &  4.341 &  1 &  1 &  1.30 &  HD95868 &  1.24  \\    
182228  & 20.2  & 3.153  & 0.170  & 1.990  & 3.209  & 2.240    &  4.422 &  1 &  1 &  1.09 &  HD73708 &  1.09  \\    
202360  & 20.8  & 3.153  & 0.132  & 2.455  & 3.566  & 2.579    &  2.198 &  2 &  2 &  1.33 &  HD19061 &  1.20  \\    
203061  & 20.8  & 3.151  & 0.157  & 0.326  & 3.155  & 2.167    &  2.293 &  3 &  2 &  1.30 &  HD19061 &  1.11  \\    
211009  & 19.5  & 3.152  & 0.132  & 1.859  & 2.759  & 1.782    &  4.007&  4 &  2 &  1.33 &  HD95868 &  1.28  \\     
\hline
{\it Veritas} &&& &&& &&& &&&\\
\hline
5592  & 17.3  & 3.173  & 0.062  & 8.511  & 3.254  & 2.696        &  15.777 &  3 &  2 &  1.05 &  HD73708 &  1.03  \\      
7231  & 17.1  & 3.168  & 0.075  & 9.429  & 2.947  & 2.215        &  14.862 &  3 &  2 &  1.20 &  HD220764 &  1.28  \\     
8624  & 18.9  & 3.166  & 0.032  & 9.080  & 3.207  & 2.384        &  11.364 &  4 &  2 &  1.04 &  HD42160 &  1.14  \\      
62512  & 19.0  & 3.165  & 0.065  & 8.807  & 3.015  & 2.040       &  3.871 &  4 &  2 &  1.05 &  HD73708 & 1.03  \\        
82961  & 18.6  & 3.163  & 0.057  & 8.317  & 3.145  & 2.163       &  3.138 &  3 &  2 &  1.09 &  HD19061 &  1.02  \\       
169282  & 20.0  & 3.166  & 0.082  & 10.236  & 3.074  & 2.536     &  17.079 &  4 &  2 &  1.04 &  HD257880 &  1.02  \\     
215270  & 19.5  & 3.172  & 0.079  & 10.117 & 2.951  & 2.027      &  8.560 &  2 &  2 &  1.15 &  HD19061 &  1.20  \\       
\enddata
\tablecomments{Orbital elements (a, e, i) ephemerides, and observational data for each asteroid.  
Observing run numbers correspond to Run 1= Mar. 02, 2013, 
Run 2 = Oct. 29, 2013, Run 3 = Oct. 30, 2013, Run 4= Feb. 21, 2014.  $^\dag$heliocentric distance 
$^\ddag$geocentric distance $^\S$phase angle
 }
\end{deluxetable}

\begin{deluxetable}{lcrrccccl}
\tabletypesize{\footnotesize}
\tablewidth{0pt}
\tablecaption{\large \bf Subaru Asteroid Characteristics \label{tab_subaru_data}}
\tablehead{
\colhead{\bf Asteroid}   &
\colhead{\bf Hmag}& 
\colhead{\bf D$^a$} &
\colhead{\bf Slope} &
\colhead{\bf SNR$^b$} &
\colhead{\bf UV$^c$}&
\colhead{\bf 0.7~$\mu$m $^c$} & 
\colhead{\bf 0.7 ~$\mu$m Band $^d$} &
\colhead{\bf Obs Run$^e$}  
 \\
\colhead{\bf ID} &
\colhead{}&
\colhead{\bf (km) } &
\colhead{\bf (\%~/~1000~$\AA$)}&
\colhead{}&
\colhead{\bf Abs}&
\colhead{\bf Abs}&
\colhead{\bf Detection Limit (\%)} &
\colhead{\bf No.}
}
\startdata
\hline
{\it Themis } & &&&&&&&\\
\hline
7511		&	13.7	& 7.09 &	-0.896 $\pm$ 0.020 &   158.5 & Y & N &  0.6 & 3   \\
8697		&	13.5	& 9.16 &	-2.165 $\pm$ 0.008 &    25.0 & N & N & 4.0 & 4 \\
15572	&	13.4	& 8.12&	0.088 $\pm$ 0.004 &    83.6 &  Y & N & 1.2 & 3 \\
15772	&	13.4	& 8.82&	 -1.917 $\pm$ 0.007 &    40.4 & N & N &  2.5 & 4  \\ 
16877	&	13.6	& 9.19 & -0.065 $\pm$ 0.010 &    62.5 &  N & N & 1.6 & 4 \\
18946	&	14.4	& 6.36 &	  3.037 $\pm$ 0.005 &  37.0 &  N & N & 2.7 & 1 \\	
26633	&	14.1	& 8.06&	-0.025 $\pm$ 0.004 &    95.9 &  Y & N & 1.0 & 3 \\
28789	&	14.4	& 6.36&	-0.900 $\pm$ 0.006 &    66.0 &  N & N & 1.5 & 3 \\
31093	&	13.4	& 9.43& 	 0.293 $\pm$ 0.013 &    82.6 &  N & N & 1.2 & 4 \\
31334	&	13.6 & 10.18& 	 0.631 $\pm$ 0.019 &   101.5 &  N & N & 1.0 & 4 \\
38365	&	14.1	& 7.79&	2.209 $\pm$ 0.017 &    35.4 &  N & N & 2.8 & 1 \\
47920	&	13.4	& 11.25&	0.437 $\pm$ 0.012 &   129.7 &  Y & N & 0.8 & 3 \\
51143	&	13.4	& 9.93 &	-1.083 $\pm$ 0.007 &   79.0 &  N & N & 1.3 & 3 \\
51579	&	14.2	& 6.01&	0.807 $\pm$ 0.004 &    33.1 &  N & N & 3.0 & 3 \\
54320	&	14.4	& 7.02&	-1.072 $\pm$ 0.015 &   124.0 &  N & N & 0.8 & 3 \\
58231	&	14.0	& 7.64&	 -1.426 $\pm$ 0.012 &  126.5 &  N & N & 0.8 & 3 \\
58302	&	14.2	& 6.91&	-1.473 $\pm$ 0.004 &  71.9 &  N & N & 1.4 & 1  \\
82646	&	14.2	& 6.97 &	  0.115 $\pm$ 0.007 &  50.5 &  N & N & 2.0 & 4 \\
91965	&	14.1 & 6.47&	  0.250 $\pm$ 0.005 &  55.2 &  N & N & 1.8 & 1 \\
123646	&	14.6	& 5.80&	-2.557 $\pm$ 0.005 & 45.8 &  N & N & 2.2 & 1 \\
132845	&	15.3	& 4.24&	 -0.688 $\pm$ 0.007 &  21.2 &  N & N & 4.7 & 3 \\
149084	&	15.0	& 4.78&  -1.161 $\pm$ 0.004 &  27.1 &  N & N & 3.7 & 1 \\
\hline
{\it Beagle  } & &&&&&&&\\
\hline
7968		&	15.6	& 4.87& -1.132 $\pm$ 0.004 &    13.5 &  N & N &7.4 & 3 \\
8161 	&	13.3	& 10.51&	-1.787 $\pm$ 0.004 &    99.3 &  N & N &1.0 & 2  \\
          	&& 			     &	 0.964 $\pm$ 0.010 &    31.2 &  N & N &3.2 & 4 \\
19923	&	13.8	& 8.08   & -0.128 $\pm$ 0.008 &    85.3 &  N & N &1.2 & 4 \\
21569	&	13.9 & 13.65 &	0.455 $\pm$ 0.015 &    55.5 &  N & N &1.8 & 1 \\
26140	&	14.1	& 7.83   & 0.418 $\pm$ 0.004 &    54.0 &  N & N &1.9 & 3 \\
55188	&	13.6	& 8.69 & -1.924 $\pm$ 0.002 &   32.9 &  N & N &3.0 & 2 \\
63951	&	15.0 & 5.54 & -1.664 $\pm$ 0.003 &    39.6 &  N & N &2.5 & 3 \\
71489	&	13.8	& 8.13 & -0.242 $\pm$ 0.003 &   28.2 &  N & N &3.5 & 3  \\
84230	&	15.1	& 5.49 & -2.572 $\pm$ 0.005 &    32.2 &  N & N &3.1 & 3  \\
88079	&	14.9 & 5.49 & -3.353 $\pm$ 0.005 &    35.6 &  N & N &2.8 & 1  \\
104671	&	15.4	& 4.36 & -1.261 $\pm$ 0.003 &    55.0 &  N & N &1.8 & 2 \\
116557	&	15.2	& 4.30 & -0.810 $\pm$ 0.005 &   33.0 &  N & N &3.0 & 3 \\
140399	&	14.6 & 6.39 &  -1.555 $\pm$ 0.003 &    40.7 & N & N &  2.5 & 3 \\
		&&&	 -2.340 $\pm$ 0.012  &  55.8 &  N & 3.4 $\pm$ 1.3 & 1.8 & 4 \\
143633	&	15.7	& 3.53& -0.423 $\pm$ 0.011 &    14.9 &  N & N & 6.7 & 4 \\
144732    &	15.6 & 3.70 & -1.082 $\pm$ 0.003 &   17.9 &  N & N & 5.6 & 2 \\
150625	&	16.1 & 3.26 & -2.444 $\pm$ 0.009 &   22.0 &  N & N & 4.5 & 1 \\	
166780	&	15.7	& 4.09 & -2.506 $\pm$ 0.004 &   58.5 &  N & N & 1.7 & 2  \\
178208	&	15.6	& 4.34 & -1.895 $\pm$ 0.008 &   41.6 &  N & N & 2.4 & 4 \\
180762	&	15.5 & 3.29 & -1.379 $\pm$ 0.013 &   10.8 &  N & N & 9.3 & 1 \\
182228	&	15.6 & 4.34 & -0.710 $\pm$ 0.008 &    23.4 &  N & N & 4.3 & 1 \\
202360	&	15.6	& 3.57 & -0.469 $\pm$ 0.003 &    21.3 &  N & N & 4.7 & 2 \\
203061	&	16.3	& 3.02 & -0.363 $\pm$ 0.007 &    26.3 &  N & N & 3.8 & 3 \\
211009	&	15.7	& 3.98 & -2.058 $\pm$ 0.010 &   70.9 &  N & N & 1.4 & 4 \\
\hline
{\it Veritas } &&&&&&&&\\
\hline
5592		&	11.7	& 22.09	& -1.366 $\pm$ 0.011 &   111.4 &  N & 3.0 $\pm$ 0.7 & 0.9  & 3 \\
7231		&	12.2	& 18.33  &	 -0.588 $\pm$ 0.007 &  103.9 &  Y & 1.1 $\pm$ 0.8 & 1.0  & 3  \\
8624		&	13.6	& 9.50 & 0.167 $\pm$ 0.002 &   49.9 &  N & 1.9 $\pm$ 1.8 & 1.9 & 4 \\
62512	&	14.7 & 5.76 &	 1.211 $\pm$ 0.013 &  73.8 &  Y & 1.4 $\pm$ 1.1 & 1.3  & 4 \\ 
82961	&	14.1	& 6.93 & -2.582 $\pm$ 0.010 &  65.7 &  N & 3.0 $\pm$ 1.2 & 1.5  & 3 \\
169282	&	14.6	& 6.27 & -0.704 $\pm$ 0.006 &  26.9 &  N & N  & 3.7 & 4 \\
215270	&	15.1	& 5.34  & 1.411 $\pm$ 0.002 &  30.2 &  N & N & 3.3 & 2  \\

\enddata
\tablecomments{Spectral data of asteroids observed with Subaru telescope. 
$^a$ Diameters (D) were computed using mean family albedos when diameter measurements
were not available from WISE. $^b$ Mean signal to noise ratios, measured
using data between 0.49 $< \lambda < 0.91~\mu$m.  
$^c$ Indication where absorption is seen in the spectrum either in the UV or at
0.7~$\mu$m.  If visible, the 0.7~$\mu$m band depth is reported. 
$^d$ Detection limits derived from the mean signal to noise ratio of each spectrum. 
$^e$ Observing runs 1, 2, 3, and 4 correspond  Mar. 02, 2013, 
Oct. 29, 2013, Oct. 30, 2013, and Feb. 21, 2014 respectively.  
 }
\end{deluxetable}

\begin{deluxetable}{lrccccc}
\tabletypesize{\footnotesize}
\tablewidth{0pt}
\tablecaption{\large \bf Asteroid Family Characteristics \label{tab_family_summary}}
\tablehead{
\colhead{\bf Family}   &
\colhead{\bf Slope} &
\colhead{\bf $\sigma_{slope}$} &
\colhead{\bf Albedo} & 
\colhead{\bf $\sigma_{albedo}$} &
\colhead{\bf Albedo} &
\colhead{\bf $\sigma_{albedo}$}  \\
\colhead{} &
\colhead{\bf \%/1000 $\AA$} &
\colhead{ } &
\colhead{\bf ($<$ 15 km)}&
\colhead{} &
\colhead{\bf ($\geq$ 15 km)} &
\colhead{}
}
\startdata
Beagle & -1.280 $\pm$ 0.003 & 1.080 & 0.0794 $\pm$ 0.0045 & 0.0179 & -- & -- \\
Themis & -0.378 $\pm$ 0.003 & 1.342 &  0.0680 $\pm$ 0.0006 & 0.0236 & 0.0747 $\pm$ 0.0013 & 0.0214 \\
Veritas & 0.196 $\pm$ 0.001 & 1.532 & 0.0695 $\pm$ 0.0010 & 0.0245 & 0.0619 $\pm$ 0.0064 & 0.0255 
\enddata
\tablecomments{
 Slopes represent weighted means for each family, and albedo means were derived from reported 
 WISE values \citep{Masiero:2011}.  The uncertainty reported with the means are the standard error 
 of the mean.  The sigma values represent the standard deviation and 
 show the dispersion for each population.
 }
\end{deluxetable}

\end{document}